\documentclass[fleqn,usenatbib]{mnras}

\usepackage{newtxtext,newtxmath}

\usepackage[T1]{fontenc}
\usepackage{ae,aecompl}

\usepackage{graphicx}	
\usepackage{amsmath}	
\usepackage{amssymb}	
\usepackage{multirow}

\title[Profiles for realistic tomographic error]{Identifying optical turbulence profiles for realistic tomographic error in adaptive optics} 

\author[O. J. D. Farley et al.]{
O. J. D. Farley,$^{1}$\thanks{E-mail: o.j.d.farley@durham.ac.uk}
J. Osborn,$^{1}$
T. Morris,$^{1}$
T. Fusco,$^{2,3}$
B. Neichel,$^{2}$
C. Correia$^{2}$
\newauthor and R. W. Wilson$^{1}$
\\
$^{1}$Centre for Advanced Instrumentation (CfAI), Dept. Physics, Durham University, Durham, DH1 3LE, UK\\
$^{2}$Aix Marseille Univ, CNRS, CNES, LAM, Marseille, France\\
$^{3}$ONERA, 29 avenue de la Division Leclerc, 92322 Ch\^{a}tillon, France\\
}

\date{Accepted XXX. Received YYY; in original form ZZZ}

\pubyear{2019}

\begin{document}
\label{firstpage}
\pagerange{\pageref{firstpage}--\pageref{lastpage}}
\maketitle

\begin{abstract}
For extremely large telescopes, adaptive optics will be required to correct the Earth's turbulent atmosphere. The performance of tomographic adaptive optics is strongly dependent on the vertical distribution (profile) of this turbulence. An important way in which this manifests is the tomographic error, arising from imperfect measurement and reconstruction of the turbulent phase at altitude. Conventionally, a small number of reference profiles are used to obtain this error in simulation however these profiles are not constructed to be representative in terms of tomographic error. It is therefore unknown whether these simulations are providing realistic performance estimates. Here, we employ analytical adaptive optics simulation that drastically reduces computation times to compute tomographic error for 10 691 measurements of the turbulence profile gathered by the Stereo-SCIDAR instrument at ESO Paranal. We assess for the first time the impact of the profile on tomographic error in a statistical manner. We find, in agreement with previous work, that the tomographic error is most directly linked with the distribution of turbulence into discrete, stratified layers. Reference profiles are found to provide mostly higher tomographic error than expected, which we attribute to the fact that these profiles are primarily composed of averages of many measurements resulting in unrealistic, continuous distributions of turbulence. We propose that a representative profile should be defined with respect to a particular system, and that as such simulations with a large statistical sample of profiles must be an important step in the design process.

\end{abstract}




\section{Introduction}

Tomographic adaptive optics (AO) techniques are key to unlocking the imaging and spectroscopic performance of extremely large telescopes (ELTs). These techniques allow the correction of the Earth's atmosphere over a large fraction of the sky as in the case of laser tomographic AO \citep[LTAO; see e.g.][]{Fusco2010a}, over a wide field of view in the case of multi-conjugate AO \citep[MCAO;][]{Beckers1988} and along multiple lines of sight in the case of multi-object AO \citep[MOAO;][]{Assemat2007}. As such these systems are fundamental to ELT operation. 


The optical turbulence profile, usually parameterised in terms of the refractive index structure constant $C_n^2 (h)$ as a function of altitude $h$, is an important parameter in computing the tomographic reconstruction of the turbulent phase. The profile is known to change on timescales of minutes to seasons and is unique to a particular location. The shape of the profile will therefore impact directly on the performance of tomographic AO systems and as a result measurements of the turbulence profile form an important part of astronomical site characterisation campaigns. 

One important source of error as a result of the profile is the tomographic error, arising from imperfect reconstruction of the turbulent phase at altitude. This error forms a large part of the error budget for many tomographic AO systems and is common to all systems for a given guide star asterism and turbulence profile regardless of particular AO flavour (LTAO, MCAO, MOAO, etc.). It is therefore important that the turbulence profiles used in simulation provide tomographic errors similar to those that would be seen on sky. The impact of the turbulence profile on the tomographic error has been modelled by \cite{Tokovinin2001}, showing that the degree to which the turbulence is confined to thin layers is of greater importance for smaller error than the overall strength of the turbulence. This is supported in simulation by \cite{Fusco2010} using a limited set of turbulence profiles.

For end to end (E2E) Monte Carlo simulations of tomographic AO, a single or very few turbulence profiles are used due to long computation times. These profiles are usually chosen by taking mean or median profiles from large databases obtained by site characterisation and monitoring campaigns \citep[see e.g.][]{Travouillon2009, DaliAli2010, Vernin2011, Osborn2018a}. Typically profiles are selected from these databases that are close to some values of integrated parameters of the turbulence such as seeing or isoplanatic angle, for which the distribution for the entire database can be easily computed \citep{Sarazin2013, Travouillon2009}. Since tomographic error depends on some parameters of the AO system in question (e.g. number of guide stars and their asterism) these profiles are not constructed with this error in mind -- instead it must be assumed a priori that, for example, a median $C_n^2$ profile will result in median tomographic AO performance.

One such large database of turbulence profiles has been produced by the Stereo-SCIDAR (SCIntillation Detection And Ranging) at ESO Paranal, Chile \citep{Osborn2018a}. This high resolution, high sensitivity turbulence profiler has been in operation since 2016 producing over 10 000 full atmosphere profiles with 250 m altitude resolution. Previous work has attempted to reduce this dataset to small sets of profiles that could be used in E2E simulations \citep{Sarazin2017, Farley2018}. However, without some form of simulation it is not possible to know whether these profiles represent the dataset in terms of tomographic error, arguably one of the most important error terms for next generation tomographic AO systems.

Here we employ a fast analytical AO simulation \citep{Neichel2008} in order to directly ascertain the impact that the varying turbulence profile has on AO performance allowing computation of the tomographic error using a large number of turbulence profiles. By operating in the Fourier domain and computing the power spectral density of the AO-corrected PSF we can obtain the tomographic error of an ELT scale AO system for a particular set of atmospheric conditions in seconds as opposed to hours on modest hardware. We are therefore not limited in the number of turbulence profiles we can consider allowing us to understand how the tomographic error is distributed over a large database of real turbulence profiles. Additionally, through comparison of the distribution of tomographic error across the Stereo-SCIDAR dataset to small sets of profiles commonly used to represent the Paranal turbulence profile in E2E simulation we can assess how representative these profiles are.


\section{Fourier simulation}
A limitation of conventional Monte Carlo AO simulation is that of convergence: the random nature of atmospheric turbulence requires many realisations of the turbulent phase to be simulated in order for the results to converge. While this allows for many complex aspects of the AO system to be simulated with high accuracy, these simulations require long computation times.

The starting point of the Fourier approach is to assume that the whole problem (phase propagation, wavefront sensor (WFS) measurements, deformable mirror (DM) commands etc.) is linear and spatially shift-invariant. In that case, all the usual operators are diagonal with respect to spatial frequencies and simply act as spatial filters in the Fourier domain. It follows that each equation can be written frequency by frequency, and that tomographic reconstruction algorithms may be derived and evaluated one Fourier component at a time. In addition, second order statistics of the residual phase and long exposure PSFs can be evaluated directly, without requiring any iterations. By avoiding the convergence problem simulation times are cut down by orders of magnitude at ELT scales from hours to seconds for a single profile on modest hardware. The main limitation of the Fourier approach is that aperture-edge effects and boundary conditions, which cannot be represented by shift-invariant spatial filters, are neglected. Hence, the Fourier modelling only applies on the idealised case of infinite aperture system, and all effects of incomplete beam overlap in the upper atmospheric layers are neglected. However, in the frame of ELTs, the size of the telescope aperture are large enough to satisfy this assumption. For instance, \cite{Neichel2008} show that the Fourier approach provides similar results as E2E simulations for a 40m telescope, as long as the guide star constellation remains smaller than a field of view diameter of 10 arcminutes. Aware of these limitations, all the following simulations have been performed well within the valid regime of the Fourier approach.

\subsection{Tomographic error vs. other errors}
For a classical single conjugate AO system, residual phase variance $\sigma_{\phi}^2$ after AO correction can be considered a sum of independent error terms
\begin{equation}
    \sigma_{\phi}^2 = \sigma_{\mathrm{fitting}}^2 + \sigma_{\mathrm{aniso}}^2 + ... \, ,
    \label{eq:SCAO_error}
\end{equation}
where we have shown here only the DM fitting error $\sigma_{\mathrm{fitting}}$ and anisoplanatic error $\sigma_{\mathrm{aniso}}$, two error terms associated with the turbulence profile.

These error terms can be modelled using the well known Fried parameter $r_0$ \citep{Fried1966} and isoplanatic angle $\theta_0$ \citep{Fried1976}:
\begin{equation}
r_0 = \left(0.423 k^2 \cos(\gamma)^{-1} \int_0^{\infty} C_n^2(h) \, \mathrm{d}h\right)^{-3/5}
\label{eq:r0},
\end{equation}
\begin{equation}
\theta_0 = \left(2.91 k^2 \cos{(\gamma)}^{-8/3} \int_0^{\infty} C_n^2 (h) h^{5/3} \, \mathrm{d}h \right)^{-3/5} = 0.314 \frac{r_0}{\bar{h}},
\label{eq:theta0}
\end{equation}
where we have $k=2\pi/\lambda$ the wavevector of light considered and $\gamma$ the zenith angle. The quantity $\bar{h}$ is the mean effective turbulence height, defined as:
\begin{equation}
    \label{eq:heff}
    \bar{h} = \left( \frac{\int_0^\infty C_n^2 (h) h^{5/3} \, \mathrm{d}h}{\int_0^\infty C_n^2 (h) \, \mathrm{d}h} \right) ^{3/5}.
\end{equation}
Note that for all calculations we use the wavelength $\lambda=500$ nm and zenith angle $\gamma=0$ where applicable.

Using these parameters we can model classical fitting and anisoplanatic error will the well known scaling laws:
\begin{equation}
    \sigma_{\mathrm{fitting}}^2 \propto \left(\frac{d}{r_0}\right)^{5/3},
    \label{eq:fitting_error}
\end{equation}
\begin{equation}
    \sigma_{\mathrm{aniso}}^2 \propto \left(\frac{\theta}{\theta_0}\right)^{5/3},
    \label{eq:aniso_error}
\end{equation}
with $d$ DM actuator pitch and $\theta$ the angular distance from the compensated direction in the field of view \citep{Rigaut1998,Fried1982}. 


The parameters $r_0$ and $\theta_0$ are simple to calculate from a turbulence profile and most profiles used in simulation are constructed such that they are representative in terms of these parameters. We know that $\sigma_{\mathrm{fitting}}$ and $\sigma_{\mathrm{aniso}}$ should be well modelled in any simulation using these profiles, and therefore the simulation will provide realistic performance of an AO system provided that the contribution of the other error terms is small.

Moving to a tomographic AO system, we are no longer able to model anisoplanatism as a purely atmospheric effect in isolation. The quality of AO correction varies across the field according to the ability of the system to reconstruct the phase aberration along any particular line of sight, which depends on the number of guide stars employed as well as the asterism configuration. This error is commonly referred to as the tomographic error. For a tomographic system, the equivalent expression to Eq. \ref{eq:SCAO_error} is
\begin{equation}
    \sigma_{\phi}^2 = \sigma_{\mathrm{tomo}}^2 + \sigma_{\mathrm{fitting}}^2 + ... \, ,
\end{equation}
where we have denoted $\sigma_{\mathrm{tomo}}$ as the tomographic error. Note that for an MCAO system, the fitting error is replaced by the generalised fitting error \citep{Rigaut2000}, which may still be modelled as an integral over the $C_n^2$ profile but is dependent on the number of DMs and their respective conjugate altitudes.

Analogous to Eqs. \ref{eq:fitting_error} and \ref{eq:aniso_error} the tomographic error for $K$ guide stars in a circle with diameter $\Theta$ may be modelled in the infinite aperture, zero noise limit as
\begin{equation}
    \sigma_{\mathrm{tomo}}^2 \propto \bigg(\frac{\Theta}{\gamma_K}\bigg)^{5/3},
    \label{eq:tomo_tokovninin}
\end{equation}
where $\gamma_K = r_0/\delta_K$ is the tomographic patch size \citep{Tokovinin2001}. Note also that a Kolmogorov turbulence spectrum is assumed here (infinite outer scale $L_0$). Unlike $r_0$ and $\theta_0$, the parameter $\delta_K$, which may be interpreted as the effective thickness of the profile for $K$ guide stars, is not computable as an integral over the profile as for $r_0$ and $\theta_0$. We therefore require a more in-depth Fourier analysis of the system to obtain the tomographic error and hence general reference profiles cannot be constructed explicitly to be representative in terms of tomographic error.



\subsection{Simulation parameters}

We simulate an ELT-like system, with the fixed simulation parameters summarised in Table \ref{tab:sim_params}. We adopt a simple tomographic AO configuration with a single DM conjugated to the ground and 6 laser guide stars (LGS) in a circular asterism. Note that while a DM is included in the simulation, we do not include any fitting error. We both optimise the tomographic reconstructor and compute the power spectral density on-axis to obtain the tomographic error. While the tomographic error will vary across the field of view, this is highly dependent on the specific parameters of the AO system therefore to maintain generality we measure only on-axis. 

\begin{table}
    \centering
    \begin{tabular}{l|r}
    Telescope Diameter & 39.3 m  \\
    Projected subaperture size & 0.5 m \\
    Projected DM pitch & 0.5 m \\
    \# LGS & 6 \\
    \# DM & 1 \\
    Tomographic reconstructor & MMSE \\
    Outer scale $L_0$ & 25 m \\
    LGS noise & 1 rad$^2$
    \end{tabular}
    \caption{Fixed simulation parameters for all LGS asterisms.}
    \label{tab:sim_params}
\end{table}

We select three LGS asterisms to cover the parameter space of possible AO systems, setting the 6 LGS in circles of 1, 2 and 4 arcminutes.

For each asterism we run the simulation through the 2018A database of turbulence profiles from the Stereo-SCIDAR at ESO Paranal. This database consists of 10 691 measurements of the turbulence profile at Paranal, collected over 83 nights between April 2016 and January 2018. We can assume that since the site of the ELT Cerro Armazones and Paranal are close in both location and altitude that the turbulence profiles will be applicable to both sites, with the condition that the surface layer turbulence may differ due to local topology. This is of negligible importance in our case since we are concerned with the tomographic error, primarily associated with high altitude turbulence.

\subsection{Comparison profiles}
We compare the distribution of tomographic error across the 10 691 profiles to some small sets of profiles that aim to be representative of different atmospheric conditions. First, the commonly used ESO 35 layer (35L) median and seeing quartile profiles \citep{Sarazin2013}. These profiles are composed of combinations of profile measurements from both Paranal and Armazones, binned by seeing and averaged. 

Second, the more recent 100 layer good, high, low and all (g/h/l/a) profiles defined from the the same Stereo-SCIDAR 2018A data set \citep{Sarazin2017}. This analysis splits the profiles into three groups: good, where the total integrated turbulence is low; high, where the profile is dominated by turbulence above the ground layer and low, where the profile is dominated by the ground layer. The mean of profiles falling into each category is taken producing three reference profiles. In addition, the ``all'' profile is the mean of the entire data set. 

Finally, the profiles produced by the clustering method described in \cite{Farley2018}. Here, cluster analysis was used to partition the 10 691 profiles into 18 groups according to their shape. Two methods of defining the centre of each cluster are used: taking the mean profile from each cluster and selecting a single profile to represent each cluster. There are therefore two sets of 18 profiles from this method.


We compute the tomographic error for these profiles with the same simulation and parameters as described above so that the resulting values are directly comparable to the distribution of all 10 691 profiles from the 2018A database.

\section{Simulation results}

\begin{figure}
    \centering
    \includegraphics[width=\columnwidth]{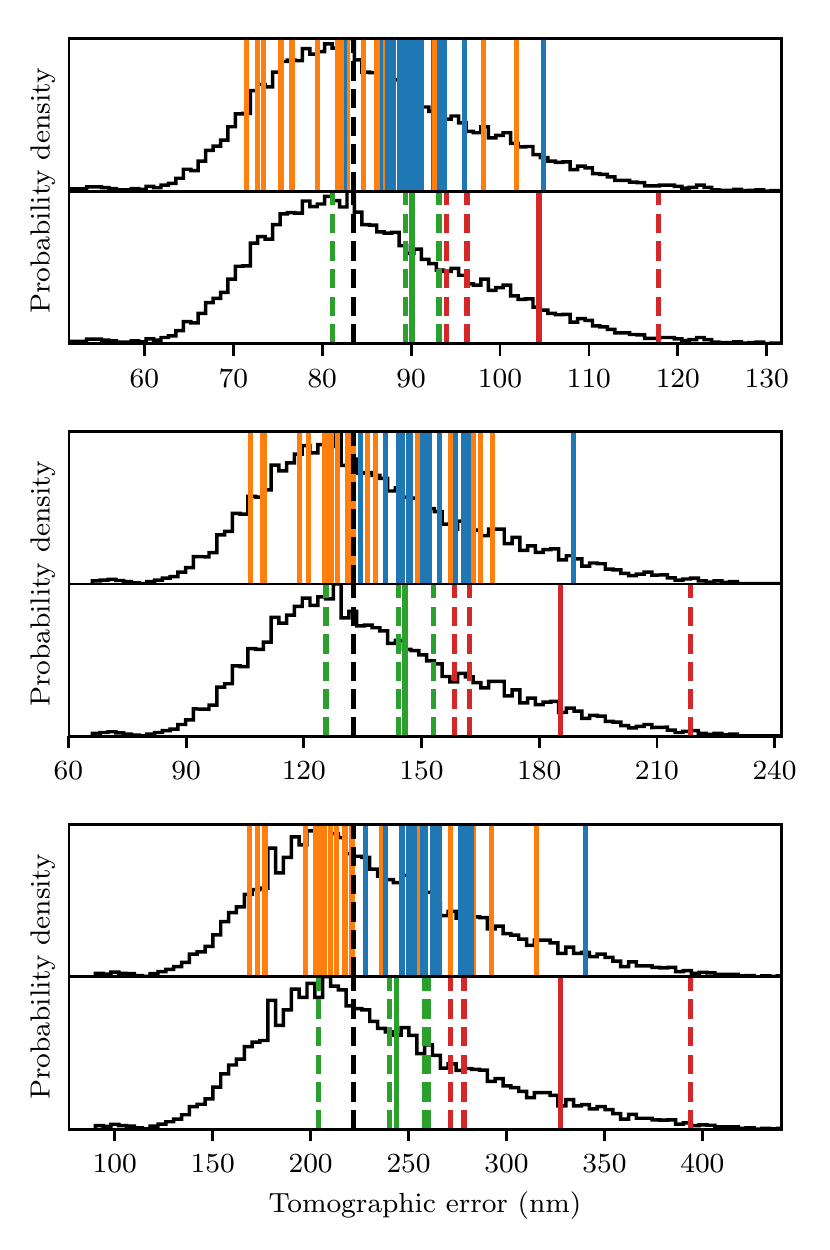}
    \includegraphics[width=\columnwidth]{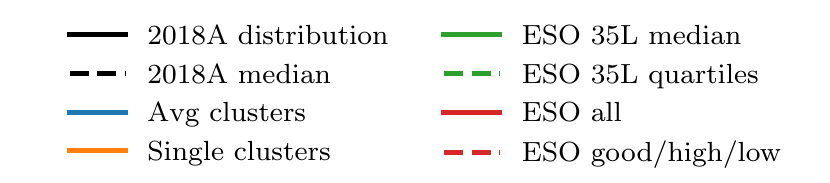}
    \caption{Histograms showing the distribution of tomographic error for all profiles (black) compared to small sets of profiles (coloured vertical lines). From upper to lower panel: 1, 2 and 4 arcminute LGS asterisms. Comparison profiles for each asterism are separated into two panels for clarity: the background histograms for each pair are identical. The median tomographic error for each asterism is indicated by the black dashed line.}
    \label{fig:histograms}
\end{figure}

The distributions of tomographic error for the simulation parameters listed in Table \ref{tab:sim_params} and our three LGS asterisms are shown in Fig. \ref{fig:histograms}. We can see that for all asterism diameters that the tomographic error across all turbulence profiles is approximately log-normally distributed, with median values of 84 nm, 133 nm and 192 nm respectively for 1, 2 and 4 arcminute asterisms. We list the values of tomographic error obtained in these simulations in Table \ref{tab:tomo_err}. 

\begin{table}
\centering
\begin{tabular}{llccc}
\hline
\bf Asterism                         &        &\bf 1' &\bf 2' &\bf 4'  \\ \hline
\multirow{3}{*}{\shortstack[l]{2018A database\\\ [10691 profiles]}}             & Q1     & 77   & 118   & 192   \\
                                   & Median & 84  & 133   & 222    \\
                                   & Q3     & 93   & 154   & 265   \\ \hline
\multirow{3}{*}{\shortstack[l]{Clusters (avg) \\\ [18 profiles]}}    & Q1     & 88   & 147  & 252     \\
                                   & Median & 91   & 151   & 260      \\
                                   & Q3     & 93   & 159   & 278      \\ \hline
\multirow{3}{*}{\shortstack[l]{Clusters (single) \\\ [18 profiles]}} & Q1     & 77 & 122   & 203   \\
                                   & Median & 83   & 130   & 215   \\
                                   & Q3     & 87   & 146   &250      \\ \hline
ESO 35L median                     &        & 90   & 146   & 244     \\ \hline
ESO 35L Q1                         &        & 81   & 126   & 204      \\ \hline
ESO 35L Q2                         &        & 89   & 144   & 240  \\ \hline
ESO 35L Q3                         &        & 93   & 153   & 258      \\ \hline
ESO 35L Q4                         &        & 93   & 153   & 260      \\ \hline
ESO good                           &        & 94   & 158   & 272      \\ \hline
ESO high                           &        & 118   & 219   & 394     \\ \hline
ESO low                            &        & 96   & 162 &  248    \\ \hline
ESO all                            &        & 104   & 185   &  327    \\ \hline
\end{tabular}
\caption{Tomographic error in nm rms for the Stereo-SCIDAR 2018A dataset as well as our comparison profiles, for each of the 1, 2 and 4 arcminute LGS asterisms simulated. We consider the 2018A dataset and the two sets of 18 clustered profiles as distributions, calculating the median, lower and upper quartiles of tomographic error for each.}
\label{tab:tomo_err}
\end{table}

Using the distributions for all the profiles of the 2018A dataset we are able to place the small sets of profiles in context. We can see that the ESO 35 layer and good/high/low/all profiles all produce tomographic errors towards the higher end of the distribution. This is also true for the clustered profiles where we take an average profile from each cluster. The clustered profiles where a single profile from each cluster is chosen to represent that cluster however are more representative of the distribution of tomographic error. 


\begin{figure}
    \centering
    \includegraphics[width=\columnwidth]{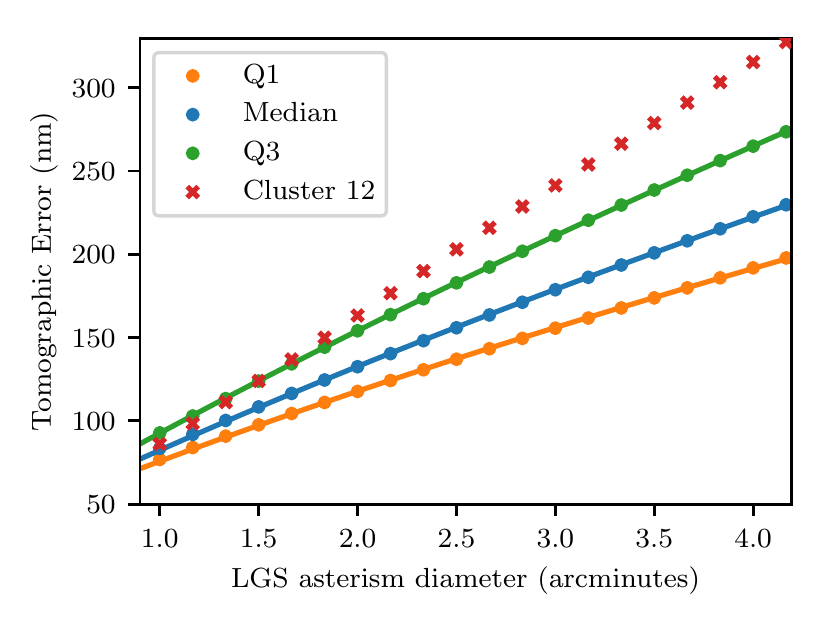}
    \caption{Computed tomographic error from Fourier simulation (circlular points and crosses) as a function of circular LGS asterism diameter. Three profiles are chosen to represent median (blue) and upper/lower quartiles (green/orange). Solid lines indicate best fit to the model in Eq. \ref{eq:tomo_fit_model}. Additionally an example of a profile whose tomographic error scaling with asterism diameter deviates substantially is shown in red crosses. Note that no fit is performed for this profile since the model does not provide a good fit for smaller asterism diameters.}
    \label{fig:ast_diam}
\end{figure}

It is informative to consider the scaling of tomographic error with LGS asterism diameter. In addition to simulations with all profiles at 1, 2 and 4 arcminutes, we select a small number of profiles with which to perform simulations with finer asterism diameter sampling. Three profiles are selected that consistently provide median, lower and upper quartile error according to the full distributions in Fig. \ref{fig:histograms}. We show how the tomographic error evolves for these profiles with asterism diameter in Fig. \ref{fig:ast_diam}. The simple 5/3 power law in Eq. \ref{eq:tomo_tokovninin} does not provide a good fit to our computed errors. We find instead an expression of the form
\begin{equation}
    \sigma_\mathrm{tomo}^2 = \alpha \Theta^{\beta} + \sigma_0^2 , 
    \label{eq:tomo_fit_model}
\end{equation}
with fit parameters $\alpha$ and $\beta$, provides a good fit to within 1\% fractional error across our asterism range. Note that the fitting is performed with error variance (nm$^2$). The parameter $\sigma_0^2$ describes the additional error resulting from non-zero noise in the LGS WFS, which may be computed by setting $\Theta=0$ in the simulation. For small asterism diameters, the error arising from the noise dominates and thus should be considered an important parameter in simulation. 

The fact that we must introduce the parameter $\beta$ in order to obtain a good fit has the important consequence that each profile scales slightly differently with asterism diameter. Therefore, for example, it is not guaranteed that a profile selected for median tomographic error at one asterism will maintain median error as the asterism changes, since it could scale more or less rapidly relative to other profiles. We show this in Fig. \ref{fig:ast_diam} with a profile (single cluster 12) that provides just over median error at 1 arcminute, but rapidly increases to around 80th percentile at 2 arcminutes and 90th percentile at 4 arcminutes. Indeed the tomographic error scales so rapidly with asterism diameter for this profile that Eq. \ref{eq:tomo_fit_model} no longer provides a good fit at small asterism diameters.

\begin{figure}
    \centering
    \includegraphics{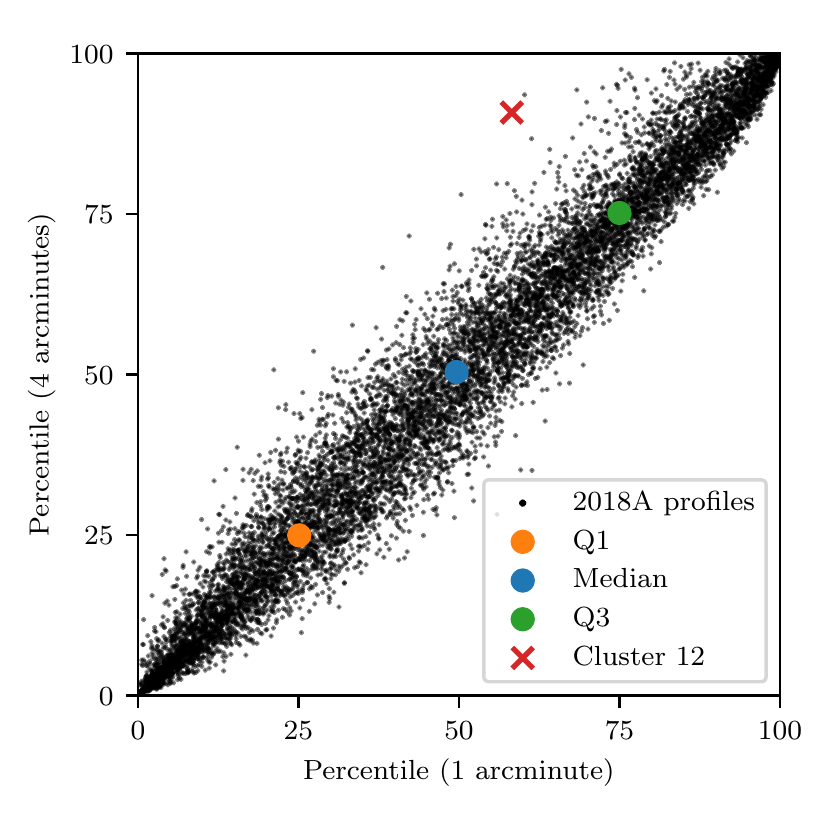}
    \caption{Percentiles of the tomographic error distribution for which the profiles of the 2018A data set (black dots) fall, for 1 arcminute and 4 arcminute asterism diameters. Also shown are the Q1, median and Q3 profiles as well as the outlier cluster 12.}
    \label{fig:percentiles}
\end{figure}

We may illustrate this over the full data set by computing the percentile of the tomographic error distribution to which each profile belongs for two asterism diameters. In Fig. \ref{fig:percentiles} we compare the percentiles of each profile in the data set for 1 arcminute and 4 arcminute diameters. We can see that the majority of profiles fall close to the $y=x$ line: as asterism diameter is increased, the profile will remain at approximately the same percentile of the distribution, with some small amount (approximately $\pm 10$ percentile) of variability. This is clearly the case for our selected median and lower/upper quartile profiles. However, there are also a clear number of outlier profiles, one of which being cluster 12, showing large shifts in terms of percentile. Care should therefore be taken when extrapolating performance between asterism diameters, as it is impossible to ascertain how the tomographic error provided by profile will scale with asterism diameter without simulation across multiple diameters and fitting to a model such as Eq. \ref{eq:tomo_fit_model}. By computing a full distribution of tomographic error as in Fig. \ref{fig:histograms} and comparing individual profiles we can be sure that we have selected a representative profile only for that particular asterism and system parameters. 

With the above caveats, from this point we will consider only the tomographic error for the 2 arcminute asterism for further analysis. We choose this particular asterism since for MCAO-like systems represented by this asterism the tomographic error is most likely to be a significant if not dominant term in the error budget. 

\begin{figure}
    \centering
    \includegraphics{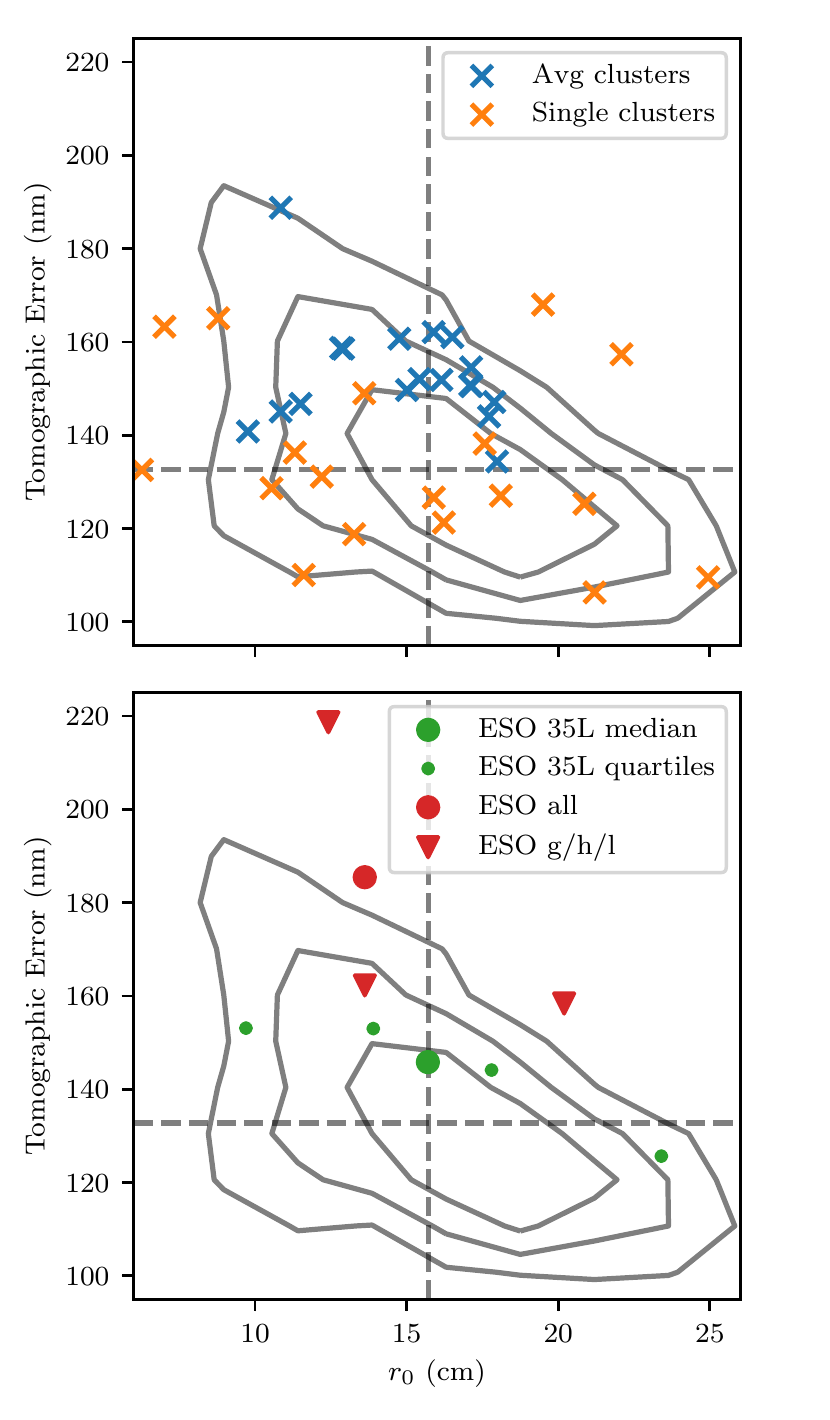}
    \caption{Two dimensional probability distributions of tomographic error for the 2 arcminute asterism against $r_0$ for clustered profiles (upper panel) and ESO profiles (lower panel). Grey contours represent the 25th, 50th and 75th percentiles of the distribution across the 2018A dataset. Dashed horizontal and vertical lines represent the median values for the considered distributions.}
    \label{fig:2d_histograms}
\end{figure}

All of the small sets of profiles considered here have been shown to be representative of the distributions of parameters $r_0$ and $\theta_0$ according to the 2018A dataset \citep{Farley2018}, with the exception of the ESO 35 layer profiles that were shown to be biased towards larger $\theta_0$. We know that the tomographic error has some dependence on $r_0$ from Eq. \ref{eq:tomo_tokovninin}. In Fig. \ref{fig:2d_histograms} we examine the two dimensional distribution of tomographic error with $r_0$ allowing us to understand how important this parameter is in determining the error.

From the contours we observe a clear correlation between $r_0$ and tomographic error, obtaining a correlation coefficient of -0.56. However we also note that the spread in tomographic error for a given value of $r_0$ is high, particularly for small values (stronger turbulence). For example, at $r_0=10$ cm, 75\% of profiles lie between around 120 nm and 200 nm tomographic error. Therefore we can confirm that, especially in strong turbulent conditions, variation in the shape of the profile is far more important than integrated strength in determining the tomographic error. Note that the correlation coefficient between turbulence effective height and tomographic error is only 0.23, indicating that this parameter is an even poorer predictor of tomographic error. 

These two dimensional distributions also show us that the additional tomographic error for profiles composed of averages (average clusters, ESO 35 layer, ESO good/high/low/all) cannot attribute this greater error to smaller values of $r_0$. These profiles follow the distribution in $r_0$ well but for any given value we obtain higher tomographic error using an average profile than we would expect. For example, the ESO 35 layer median profile has an $r_0$ value of 15.7 cm that coincides almost exactly with the median value from the Stereo-SCIDAR. However, the tomographic error for this profile is higher than expected at 146 nm; an additional 60 nm rms when added in quadrature from the median tomographic error of 133 nm. The single profile clusters are better distributed around the parameter space of $r_0$ and tomographic error and do not exhibit this bias towards higher error. There is however a noticeable lack of low $r_0$, high tomographic error profiles in this small set.

\section{Discussion}\label{sec:discussion}

We have identified two key problems regarding the choice of representative turbulence profiles. First, that while a clear correlation exists, $r_0$ is a poor predictor of tomographic error. Thus any profile or set of profiles designed to be representative in $r_0$ has no guarantee of being representative in terms of tomographic error. Second, we have found that profiles composed of averages of many measurements produce higher tomographic error than expected, given the distribution of error across the 10 691 individual profiles.

\begin{figure}
    \centering
    \includegraphics{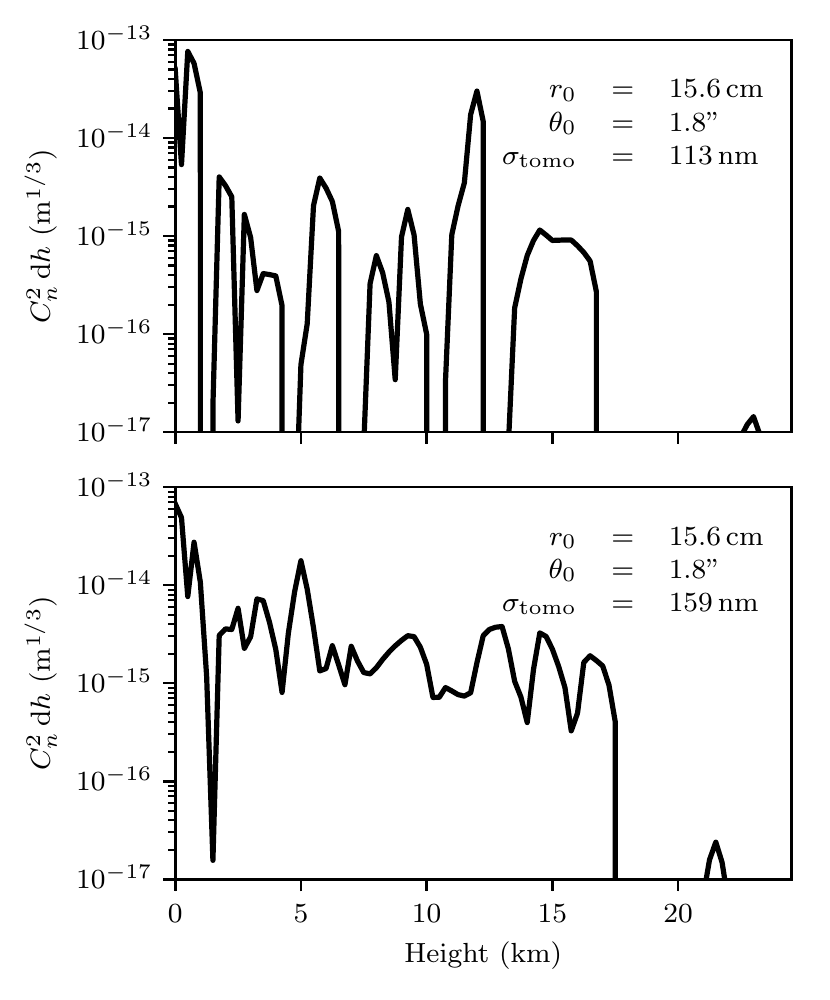}
    \caption{Two turbulence profiles taken from the Stereo-SCIDAR 2018A dataset with similar $r_0$ and $\theta_0$ values but different tomographic error, calculated here is calculated for the 2 arcminute asterism.}
    \label{fig:different_profiles}
\end{figure}

To expand on the first point we show in Fig. \ref{fig:different_profiles} two example profiles drawn from the Stereo-SCIDAR 2018A dataset. Here we select two profiles with similar $r_0$ but different tomographic errors. We also maintain similar $\theta_0$ values to ensure there is no large difference in the effective height of the turbulence. 

We can see that despite large variability in tomographic error from 113 nm to 159 nm, we are able to select two profiles that vary by less than 0.1 cm in $r_0$ and less than 0.1 arcseconds in $\theta_0$. Looking at the shape of the profiles we can contrast strong thin layers of turbulence leading to low tomographic error with a more spread out profile with weaker turbulence at all heights leading to higher tomographic error. This is consistent with previous work showing that the thickness of turbulent layers is important in the calculation of tomographic error \citep{Tokovinin2001, Fusco2010}.

\subsection{The effect of averaging profiles}
It is clear that measured turbulence profiles with less stratified, more continuous distributions of turbulence result in higher tomographic error. This characteristic of the profile is shared by profiles that are constructed by averaging a large number of measurements---over a large sample we observe both strong and weak turbulence at all altitudes across different individual profiles, therefore by taking an average profile we obtain a continuous distribution with some moderate level of turbulence at all heights. It is possible therefore that the higher tomographic error provided by reference turbulence profiles is a direct consequence of the averaging method employed to compute them.

Here, we quantify the effect of averaging on a small set of profiles by investigating the impact of averaging on tomographic error as well as the parameters $r_0$ and $\theta_0$. We select 50 profiles from the 2018A database that lie closest to the median values of $r_0$, $\theta_0$ and tomographic error concurrently. From this set of 50 individual profiles we compute mean $\overline{C_n^2} (h,N)$ and median $\widetilde{C_n^2} (h,N)$ profiles:
\begin{equation}
    \label{eq:mean_profile}
    \overline{C_n^2} (h,N) = \frac{1}{N} \sum_{i=1}^N C_{n}^2 (h,i),
\end{equation}
\begin{equation}
    \label{eq:median_profile}
    \widetilde{C_n^2} (h,N) = \mathrm{median}\left(\frac{C_n^2 (h, i \leq N)}{\int C_n^2 (h,i \leq N) \, \mathrm{d}h}\right) (C_n^2 \, \mathrm{d}h)_{\mathrm{ref}}.
\end{equation}
where $N$ is the total number of profiles used in the average computation. Thus, for example, $\overline{C_n^2} (h,25)$ represents the mean of the first 25 profiles in the 50 profile set. Note that in the case of median profiles, we must first divide the profile by its integrated $C_n^2$ then fix the integrated strength to some reference value $(C_n^2 \, \mathrm{d}h)_{\mathrm{ref}}$. The normalisation step must be taken as the median profile, unlike the mean profile, does not preserve the total turbulence sum. This is a consequence of the log-normal distribution of $C_n^2$ in each altitude bin: for this distribution the median value is always less than the mean value hence the turbulence sum always decreases, even if all profiles are normalised to the same seeing. The result is a rapid increase in $r_0$ and $\theta_0$ as $N$ increases. We choose $(C_n^2 \, \mathrm{d}h)_{\mathrm{ref}} = 3.3 \times 10^{-13}$ m$^{1/3}$, the median integrated $C_n^2$ for the 2018A data set. This means that by definition the $r_0$ values of median profiles will be constant with increasing $N$.



\begin{figure}
    \centering
    \includegraphics{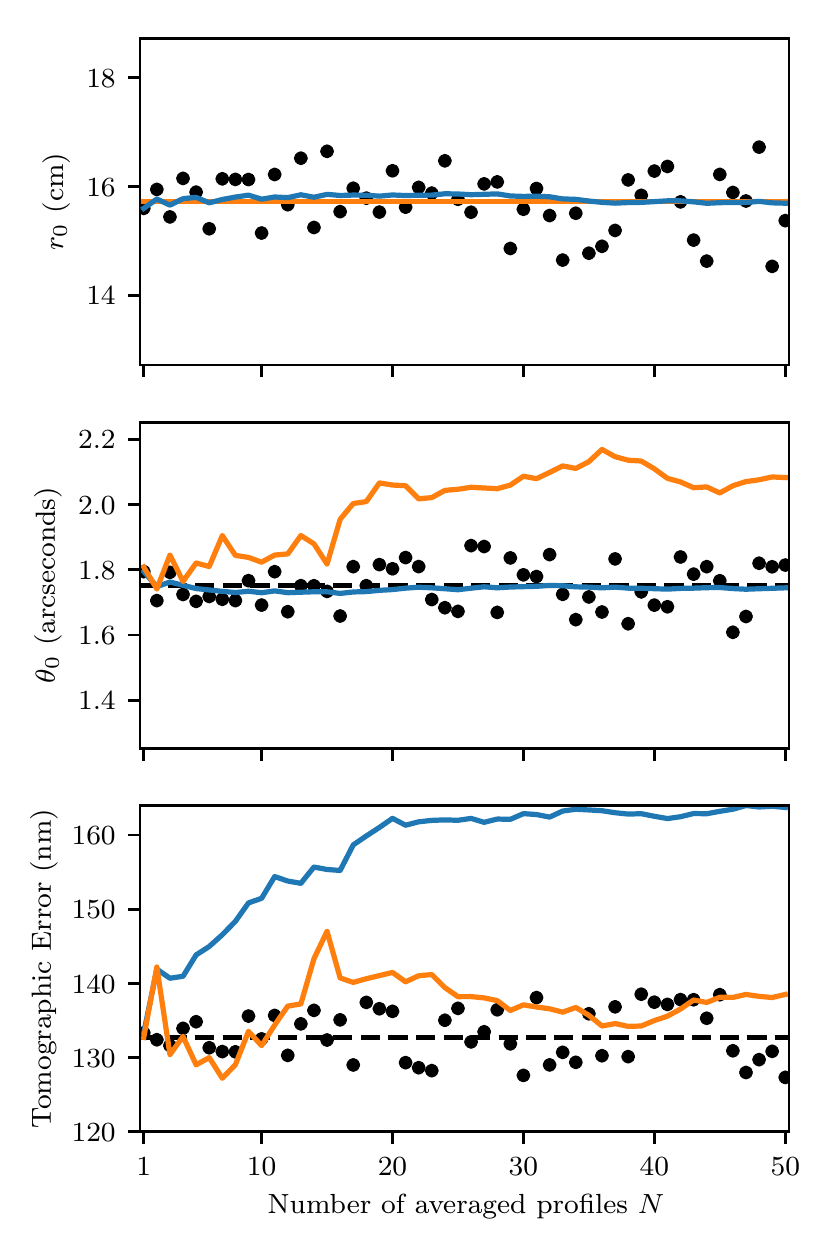}
    \caption{Evolution of $r_0$, $\theta_0$ and tomographic error as greater numbers of profiles are averaged. The values for each individual profile $i=$ 1 to 50 are indicated by black circles, with the values for mean $\overline{C_n^2} (h,N)$ and median $\widetilde{C_n^2} (h,N)$ profiles shown as blue and orange solid lines respectively. The dashed black horizontal lines indicate the median values of the respective distributions.}
    \label{fig:wfe_avgs}
\end{figure}

\begin{figure}
    \centering
    \includegraphics{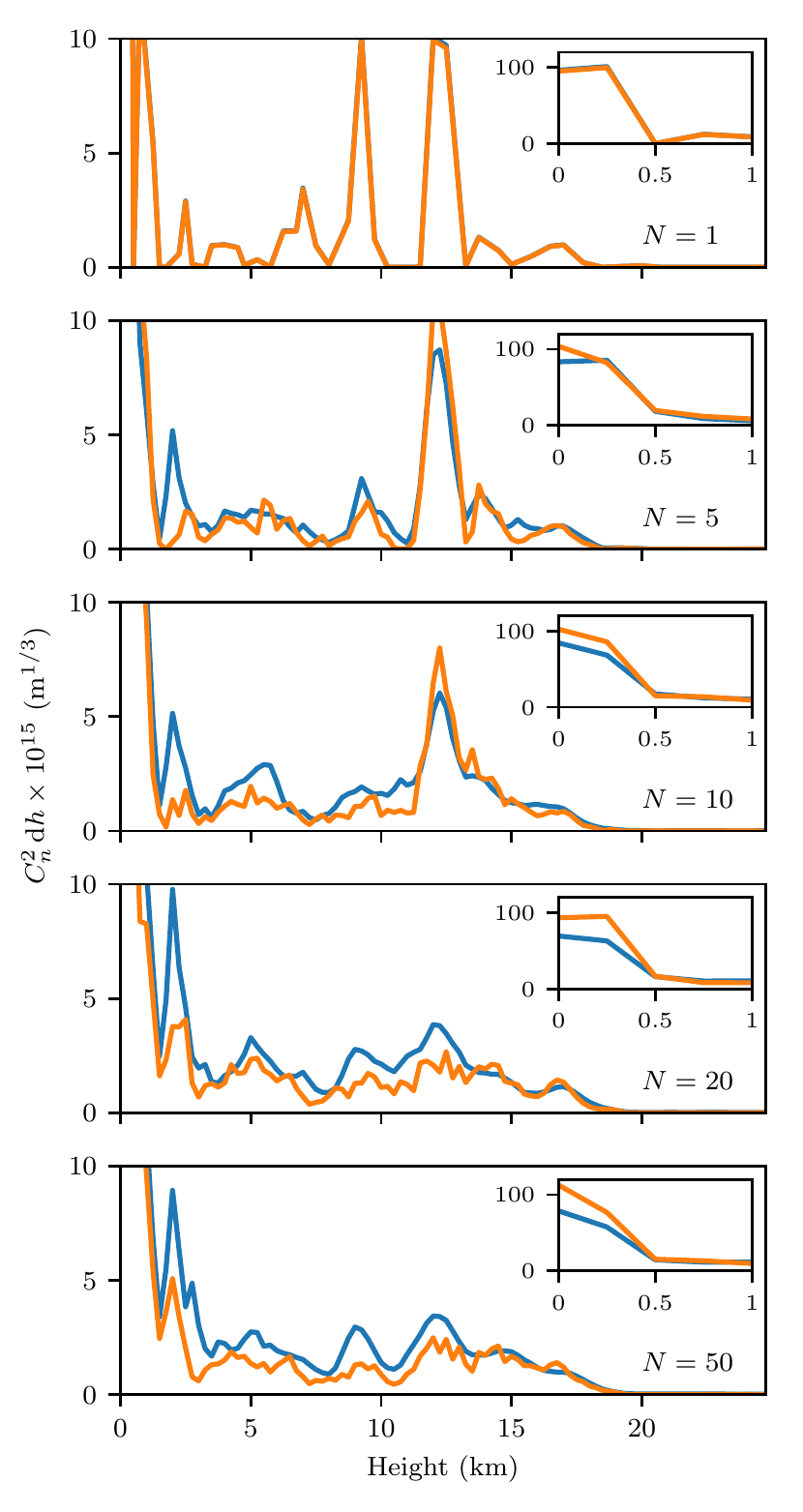}
    \caption{Turbulence profiles composed of averages of increasing numbers of profiles. From the upper to lower panel we show a single profile then the averages of 5, 10, 20, and 50 profiles. Mean $\overline{C_n^2} (h,N)$ and median $\widetilde{C_n^2} (h,N)$ profiles are shown as blue and orange lines respectively. Note that the y-axis scale is contracted to highlight the difference between profiles in the free atmosphere. Ground layer ($h<1$ km) evolution for both mean and median profiles is shown in the inset axes. At $N=1$ the two profiles are identical.}
    \label{fig:avg_profiles}
\end{figure}

In Fig. \ref{fig:wfe_avgs} we show how $r_0$, $\theta_0$ and tomographic error evolve as $N$ increases, i.e. as more profiles are included in our averaging. In terms of $r_0$ and $\theta_0$, it is clear that a mean profile represents the profiles used to compute it well, since the values of these parameters converge to their respective median values. However, the mean profile produces a much higher tomographic error than any of its constituent profiles. After averaging only 20 profiles, tomographic error has increased from 133 nm to over 160 nm rms. 

In contrast median profiles, which by design maintain a constant $r_0$ with increasing $N$, do not show as great an increase in tomographic error, reaching only around 140 nm at $N=50$. However, we now see an increase in $\theta_0$, from 1.8 arcseconds at $N=1$ to 2.1 arcseconds at $N=50$. 

To understand this increase it is informative to compare the profiles produced by our averaging methods. In Fig. \ref{fig:avg_profiles} we show the average profiles produced with our mean and median methods for 5 values of $N$. It is clear in both cases we move from strong, discrete turbulent layers to weak turbulence spread over all altitudes as $N$ increases. However, there is a marked difference here between mean and median profiles: in taking a median profile we reduce turbulence strength in the free atmosphere and increase the fraction of turbulence in the ground layer compared to the mean profile. Indeed, at $N=50$, we obtain a ground layer ($h<1$ km) fraction of 62\% by taking the median. This ground layer fraction falls at the 90th percentile of the distribution of ground layer fractions for the 50 constituent profiles. We are therefore producing a profile with a higher ground layer fraction than expected from an average profile, which in turn results in a larger $\theta_0$.

It should be noted that, despite this high ground layer fraction, the tomographic error for the $N=50$ median profile is still higher than almost all the constituent profiles. This is a result of the smoothing effect of the averaging process.

It is clear that for both averaging methods turbulence that is confined to stratified, strong layers in each individual constituent profile is spread over a wide altitude range. While in the case of the mean we can still maintain reasonable values of atmospheric parameters $r_0$ and $\theta_0$, the effect on tomographic error is large and therefore this profile will not provide realistic performance estimates if used in simulation. For the median profile, the effect on tomographic error is smaller, but this is due to an unrealistically high ground layer fraction which may effect other aspects of the simulation such as anisoplanatism.

\subsection{Choosing a representative turbulence profile}

By averaging a number of measurements of the turbulence profile, we produce an unrealistic profile that is not representative in terms of tomographic error and, in the case of median profiles, $\theta_0$. We therefore propose that when presented with a large database of turbulence profiles, one should select single profiles with the desired characteristics for the particular system in question. For instance, if a profile representing ``median" atmospheric conditions is required, a profile should be selected that lies at the median of the distribution of tomographic error as well as the median of the distributions of $r_0$ and $\theta_0$ to ensure that other errors (e.g. fitting error) are also at their respective median values. This of course requires one to compute the distribution of tomographic error for a particular instrument over the large database, but we have shown here that this may be accomplished in a feasible timescale using analytical AO simulation. 

\subsection{Extrapolation to fewer turbulent layers}

\begin{figure}
    \centering
    \includegraphics{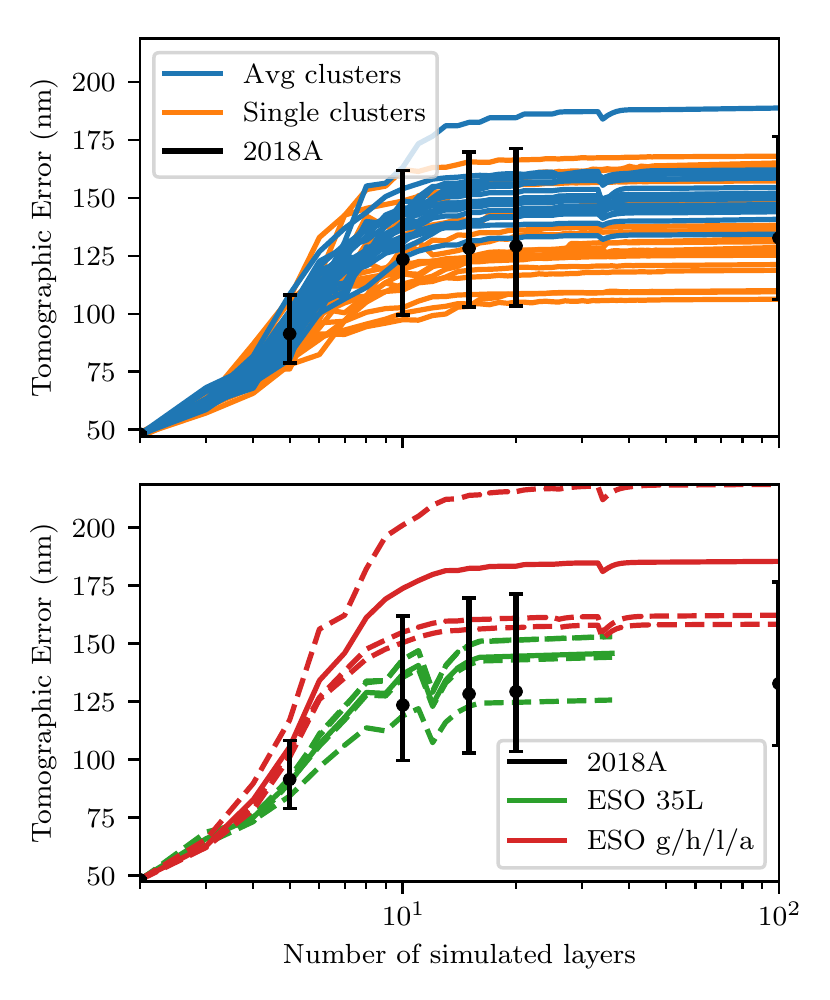}
    \caption{Evolution of tomographic error in the 2 arcminute LGS asterism case with the number of layers simulated. Error bars indicate the 10th and 90th percentiles of the tomographic error distribution over the 2018A dataset. Note that some instabilities in the equivalent layers compression method result in variability of the tomographic error, as is most apparent in the ESO 35 layer profiles. These instabilities only affect specific numbers of layers and can therefore be ignored.}
    \label{fig:nlayers}
\end{figure}

Typically in Monte Carlo simulation high resolution turbulence profiles with many turbulent layers are not used since each additional simulated turbulent layer adds to the computational complexity. It has been shown that between 10 and 20 layers are required to avoid an underestimation of tomographic error by undersampling of the profile \citep{Fusco2010}, depending on the particular system and method used to compress the profile \citep{Saxenhuber2017}. In order for our conclusions about tomographic error to be valid in Monte Carlo simulation we therefore require that there is no significant change in error as we reduce the number of layers. In Fig. \ref{fig:nlayers} we show how the tomographic error depends on the number of simulated layers for our comparison profiles and the distribution over all profiles. To compress the profiles we use the equivalent layers method \citep{Fusco1999}, which reduces the number of layers while maintaining the same $r_0$ and $\theta_0$. We can see that in all cases for very few layers the tomographic error is greatly underestimated, and the variability in tomographic error between profiles is also greatly reduced. This is understandable as it is more difficult to model a complex turbulence profile with fewer layers hence the data set becomes more homogeneous. As we increase the number of simulated layers, we see that the tomographic error begins to converge at around 10 layers, consistent with the findings of \cite{Fusco2010}. We therefore require at least 10, preferably 20 layers to sufficiently sample the turbulence profile for a 2 arcminute LGS asterism. Once we are above this threshold, the distributions of tomographic error over all of the profiles simulated are almost identical to the full 100 layer cases, and as such our conclusions drawn with 100 layer profiles are applicable to more realistic numbers of layers for Monte Carlo simulation.


\section{Conclusions}

By employing fast analytical Fourier simulation of a simple ELT-scale tomographic AO system and a large database of Stereo-SCIDAR turbulence profiles from ESO Paranal, we have performed a statistical study of tomographic error over a wide range of circular 6 LGS asterisms.

We find that the tomographic error across the whole data set follows an approximately log-normal distribution, with the median and spread increasing with increasing asterism diameter. However, further analysis shows that the tomographic error scales differently with asterism diameter for each individual profile in the data set. This means that a profile providing e.g. median error for one asterism may not necessarily be considered a median profile for other asterisms.

Our findings are consistent with previous work regarding the influence of $r_0$ on the tomographic error --- there is a moderate negative correlation between $r_0$ and tomographic error however particularly for small values the spread in computed values of tomographic error can be very large. Therefore especially in strong turbulence conditions the shape of the profile becomes of primary importance in determining the tomographic error. We observe that profiles with strong, thin layers lead to smaller error whilst more continuous distributions of turbulence lead to higher tomographic error. Additionally we find that the effective turbulence height $\bar{h}$ and therefore $\theta_0$ are not well correlated with tomographic error.

After computation of the distribution of tomographic error across over 10 691 profiles, we are able compare to small sets of profiles used in E2E simulation. We find that the ESO 35 layer, good/high/low/all and average clustered profiles are all pessimistic in terms of tomographic error to different extents. Taking as an example the 2 arcminute asterism, we find that the ESO 35 layer median and ESO all profiles result in around 60 and 130 nm additional rms error respectively when compared to the median error of the distribution of 10 691 profiles. Single profile clusters did not exhibit this bias toward higher error and perform best in terms of representing the distribution of tomographic error.

We note that the small sets of profiles providing higher tomographic error than expected are composed of averages of many measurements of the profile. By selecting 50 profiles from the dataset with similar $r_0$, $\theta_0$ and tomographic error values and cumulatively averaging them we observe a transition from a profile with strong thin layers to a continuous average profile associated with high tomographic error. We find that mean profiles are representative of their constituent profiles in $r_0$ and $\theta_0$ but rapidly increase in tomographic error. Median profiles do not increase as much in tomographic error but produce a much larger ground layer fraction than their constituent profiles and therefore larger, unrepresentative $\theta_0$.

Finally it was shown that the distributions of tomographic error computed for high resolution 100 layer profiles do not change when the number of layers simulated is reduced to more reasonable numbers for Monte Carlo simulation (fewer than 35 layers). We do not find any significant change of the tomographic error distribution for greater than 10 layers in the 2 arcminute LGS diameter case.

We propose that if profiles that are representative in terms of tomographic error are required from a large database, that single profiles should be selected from this database rather than average (mean or median) profiles. These single profiles should be selected such that they exhibit the desired error characteristics (e.g. good, median or bad tomographic error) for the given system parameters, including the guide star asterism. Fast analytical AO simulation such as has been employed here is therefore required in order to compute the distribution of tomographic error for this system across the large database such that these single profiles may be selected. Selecting a single profile in this way ensures that the characteristics of the turbulence profile as seen by the AO system are preserved, and thus any more detailed E2E simulations using this profile will produce realistic performance estimates.

\section*{Acknowledgements}

This work was supported by the Science and Technology Funding Council (UK) (ST/P000541/1) and (ST/PN002660/1). OJDF acknowledges the support of STFC (ST/N50404X/1).

Horizon 2020: This project has received funding from the European Union's Horizon 2020 research and innovation programme under grant agreement No 730890. This material reflects only the authors views and the Commission is not liable for any use that may be made of the information contained therein.

CC received support from A*MIDEX (project no. ANR-11- IDEX-0001- 02) funded by the ``Investissements d'Avenir" French Government program, managed by the French National Research Agency (ANR).

This work was supported by the Action Sp\'ecifique Haute R\'esolution Angulaire (ASHRA) of CNRS/INSU co-funded by CNES.

This research made use of Python including NumPy and SciPy \citep{VanderWalt2011}, Matplotlib \citep{Hunter2007} and Astropy, a community-developed core Python package for Astronomy \citep{Robitaille2013}. We also made use of the Python AO utility library AOtools (\url{https://github. com/AOtools/aotools}).




\bibliographystyle{mnras}
\bibliography{library} 








\bsp	
\label{lastpage}
\end{document}